\documentclass[aps,tightenlines]{revtex4}%
\usepackage{amsfonts}
\usepackage{amsmath}
\usepackage{amssymb}
\usepackage{graphicx}%
\begin{document}
\title[The $K-$way negativities and three tangle]{Partial $K-$way negativities and three tangle for three qubit states}
\author{S. Shelly Sharma}
\email[shelly@uel.br]{}
\affiliation{Departamento de F\'{\i}sica, Universidade Estadual de Londrina, Londrina
86051-990, PR Brazil }
\author{N. K. Sharma}
\email[nsharma@uel.br]{}
\affiliation{Departamento de Matem\'{a}tica, Universidade Estadual de Londrina, Londrina
86051-990 PR, Brazil }
\thanks{}

\begin{abstract}
We obtain, analytically, the global negativity, partial $K-$way negativities
($K=2$, $3)$, Wootter's tangle and three tangle for the generic three qubit
canonical state. It is found that the product of global negativity and partial
three way negativity is equal to three tangle, while the partial two way
negativity is related to tangle of qubit pairs. We also calculate similar
quantities for the state canonical to a single parameter ($0<q<1$) pure state
which is a linear combination of a GHZ state and a W state. In this case for
$q=0.62685,$ the state has zero three tangle and zero three-way negativity,
having only W-like entanglement. The difference between the product of global
and partial three way negativity and three tangle for a given state is a
quantitative measure of two qubit coherences transformed by unitary
transformations on canonical state into three qubit coherences. The global
negativity and partial $K-$way negativities, obtained by selective partial
transpositions on multi-qubit state operator, satisfy inequalities which for
three qubits are equivalent to CKW (Coffman-Kundu-Wootter) inequality.

\end{abstract}
\maketitle


Quantifying multipartite entanglement is an important problem in quantum
information theory \cite{bouw00}. A multipartite quantum system can have more
than one type of qualitatively distinct quantum correlations since a given
subsystem may be entangled to the rest in many different ways. Consequently, a
single quantity can not characterize multipartite entanglement. The first step
towards understanding the amount and nature of entanglement available to a
given party holding part of the composite system, is to quantify the quantum
correlations present in the composite system state. Peres- Horodecki
\cite{pere96,horo96} positive partial transpose is a widely used separability
criterian for a bipartite state. Negativity \cite{zycz98} of the partial
transpose of an entangled state has been shown to be an entanglement monotone
\cite{vida02}. In a recent paper \cite{shar08}, we defined $K-$way
negativities to characterize $K-$partite quantum correlations in a
multipartite state. The specific constraints applied during the construction
of $K-$way partial transpose of a state $\widehat{\rho}=\left\vert
\Psi\right\rangle \left\langle \Psi\right\vert $ relate the partial $K-$way
negativiy to $K-$partite correlations present in the state in a very natural
way. The global partial transpose can be written as a sum of $K-$way partial
transposes. The partial $K-$way negativity, defined as the contribution of
$K-$way partial transpose to global negativity, quantifies the $K-$partite
coherences of the state. A pure state $\Psi$ can be transformed, through local
unitary operations and classical communication (LOCC), to a state $\Psi_{c}$
such that both the states perform the same tasks in quantum information
processing, however, the probabilities of success may be different. If
$\Psi_{c}$ is a superposition of minimum number of local basis states and
depends on coeficients that are non-local invariants \cite{cart99}, it is
called a canonical state. The states $\Psi$ and $\Psi_{c}$ have the same
entanglement content but different quantum coherences. It was conjectured in
\cite{shar08} that the global negativities and partial $K-$way negativities
calculated for the N$-$partite canonical state are entanglement measures. In
this letter, we show that for the case of three qubit states, the two-way and
three-way partial negativities calculated for the generic canonical states are
indeed entanglement monotones. Three tangle, introduced by coffman et al.
\cite{coff00}, is a widely used measure of GHZ like entanglement of a three
qubit state. The three tangle has been shown to be an entanglement monotone
\cite{dur00}. We show that the product of global negativity and partial
$3-$way negativity for the generic three qubit canonical state \cite{acin00}
is equal to three tangle, while the product of global negativity and partial
$2-$way negativity for a given pair of qubits equals the tangle. To further
elucidate the relation between the three tangle and partial $K-$way
negativities of the canonical state, we analyze the global and partial $K-$way
negativities ($K=2,3$) of a single parameter pure state, obtained by taking a
linear combination of a GHZ state and a W state.

\section{Partial $K-$way negativities and three tangle for generic three qubit
canonical state}

The Hilbert space associated with a quantum system composed of three
sub-systems is spanned by basis vectors of the form $\left\vert i_{1}%
i_{2}i_{3}\right\rangle ,$ where $i_{m}=0$ to $\left(  d_{m}-1\right)  ,$
$d_{m}$ being the dimension of Hilbert space associated with $m^{th}$
sub-system. We refer to the first, second and third sub-system as $A$, $B$,
and $C.$ The state operator for a general tripartite state is
\begin{equation}
\widehat{\rho}=\sum_{\substack{i_{1}i_{2}i_{3},\\j_{1}j_{2}j_{3}}}\left\langle
i_{1}i_{2}i_{3}\right\vert \widehat{\rho}\left\vert j_{1}j_{2}j_{3}%
\right\rangle \left\vert i_{1}i_{2}i_{3}\right\rangle \left\langle j_{1}%
j_{2}j_{3}\right\vert . \label{1}%
\end{equation}
The global partial transposes $\widehat{\rho}_{G}^{T_{A}}$, $\widehat{\rho
}_{G}^{T_{B}}$, and $\widehat{\rho}_{G}^{T_{C}},$ are obtained from the matrix
$\rho$ by imposing the conditions
\begin{equation}
\left\langle i_{1}i_{2}i_{3}\right\vert \widehat{\rho}_{G}^{T_{A}}\left\vert
j_{1}j_{2}j_{3}\right\rangle =\left\langle j_{1}i_{2}i_{3}\right\vert
\widehat{\rho}\left\vert i_{1}j_{2}j_{3}\right\rangle \text{,} \label{2}%
\end{equation}%
\begin{equation}
\left\langle i_{1}i_{2}i_{3}\right\vert \widehat{\rho}_{G}^{T_{B}}\left\vert
j_{1}j_{2}j_{3}\right\rangle =\left\langle i_{1}j_{2}i_{3}\right\vert
\widehat{\rho}\left\vert j_{1}i_{2}j_{3}\right\rangle \text{,}%
\end{equation}%
\begin{equation}
\left\langle i_{1}i_{2}i_{3}\right\vert \widehat{\rho}_{G}^{T_{C}}\left\vert
j_{1}j_{2}j_{3}\right\rangle =\left\langle i_{1}i_{2}j_{3}\right\vert
\widehat{\rho}\left\vert j_{1}j_{2}i_{3}\right\rangle \text{.}%
\end{equation}
Global Negativity, defined as
\begin{equation}
N_{G}^{p}=\frac{1}{d_{p}-1}\left(  \left\Vert \rho_{G}^{T_{p}}\right\Vert
_{1}-1\right)  , \label{3}%
\end{equation}
measures the entanglement of subsystem $p$ with its complement in a bipartite
split of the composite system. Here $\left\Vert \rho\right\Vert _{1}$ is the
trace norm of $\rho$. Global negativity vanishes on PPT-states and is equal to
the entropy of entanglement on maximally entangled states.

The\ $K-$way partial transpose ($K$ $=2,3$) of tri-partite state
$\widehat{\rho}$ with respect to subsystem $A$ is obtained by applying the
following constraints:
\begin{align}
\left\langle i_{1}i_{2}i_{3}\right\vert \widehat{\rho}_{K}^{T_{A}}\left\vert
j_{1}j_{2}j_{3}\right\rangle  &  =\left\langle j_{1}i_{2}i_{3}\right\vert
\widehat{\rho}\left\vert i_{1}j_{2}j_{3}\right\rangle ,\quad\text{if}\quad
\sum\limits_{m=1}^{3}(1-\delta_{i_{m},j_{m}})=K,\nonumber\\
\left\langle i_{1}i_{2}i_{3}\right\vert \widehat{\rho}_{K}^{T_{A}}\left\vert
j_{1}j_{2}j_{3}\right\rangle  &  =\left\langle i_{1}i_{2}i_{3}\right\vert
\widehat{\rho}\left\vert j_{1}j_{2}j_{3}\right\rangle \quad\text{if}\quad
\sum\limits_{m=1}^{3}(1-\delta_{i_{m},j_{m}})\neq K, \label{4}%
\end{align}
where $\delta_{i_{m},j_{m}}=1$ for $i_{m}=j_{m}$, and $\delta_{i_{m},j_{m}}=0$
for $i_{m}\neq j_{m}$. The partially transposed operators $\widehat{\rho}%
_{K}^{T_{B}}$, and $\widehat{\rho}_{K}^{T_{C}}$ are defined in analogous
fashion. The $K-$way negativity calculated from $K-$way partial transpose of
matrix $\rho$ with respect to subsystem $p$, is defined as
\begin{equation}
N_{K}^{p}=\frac{1}{d_{p}-1}\left(  \left\Vert \rho_{K}^{T_{p}}\right\Vert
_{1}-1\right)  . \label{7}%
\end{equation}

A measure of $2-$way coherences involving a given pair of subsystem can be
obtained from a $2-$way partial transpose constructed by restricting the
transposed matrix elements of $\widehat{\rho}$ to those for which the state of
the third subsystem does not change. For example, $\rho_{2}^{T_{A-AB}}$ is
obtained from $\widehat{\rho}$ by applying the condition%
\begin{align}
\left\langle i_{1}i_{2}i_{3}\right\vert \widehat{\rho}_{2}^{T_{A-AB}%
}\left\vert j_{1}j_{2}i_{3}\right\rangle  &  =\left\langle j_{1}i_{2}%
i_{3}\right\vert \widehat{\rho}\left\vert i_{1}j_{2}i_{3}\right\rangle
;\quad\text{if}\quad\sum\limits_{m=1}^{3}\left(  1-\delta_{i_{m},j_{m}%
}\right)  =2,\label{10}\\
\left\langle i_{1}i_{2}i_{3}\right\vert \widehat{\rho}_{2}^{T_{A-AB}%
}\left\vert j_{1}j_{2}j_{3}\right\rangle  &  =\left\langle i_{1}i_{2}%
i_{3}\right\vert \widehat{\rho}\left\vert j_{1}j_{2}j_{3}\right\rangle
;\quad\text{for all other matrix elements}.
\end{align}
Similarly, matrix elements of $\widehat{\rho}_{2}^{T_{A-AC}}$ are related to
matrix elements of the state operator by%
\begin{align}
\left\langle i_{1}i_{2}i_{3}\right\vert \widehat{\rho}_{2}^{T_{A-AC}%
}\left\vert j_{1}i_{2}j_{3}\right\rangle  &  =\left\langle j_{1}i_{2}%
i_{3}\right\vert \widehat{\rho}\left\vert i_{1}i_{2}j_{3}\right\rangle
;\quad\text{if}\quad\sum\limits_{m=1}^{3}\left(  1-\delta_{i_{m},j_{m}%
}\right)  =2,\text{ }\quad\nonumber\\
\left\langle i_{1}i_{2}i_{3}\right\vert \widehat{\rho}_{2}^{T_{A-AC}%
}\left\vert j_{1}j_{2}j_{3}\right\rangle  &  =\left\langle i_{1}i_{2}%
i_{3}\right\vert \widehat{\rho}\left\vert j_{1}j_{2}j_{3}\right\rangle
;\quad\text{for all other matrix elements}. \label{ab}%
\end{align}
The negativities $N_{2}^{A-AB}=\left(  \left\Vert \widehat{\rho}_{2}%
^{T_{A-AB}}\right\Vert _{1}-1\right)  $ and $N_{2}^{A-AC}=\left(  \left\Vert
\widehat{\rho}_{2}^{T_{A-AC}}\right\Vert _{1}-1\right)  ,$ measure the $2-$way
coherences involving the pairs of subsystems $AB$ and $AC$, respectively.

Any three qubit pure state may be transformed to canonical form with respect
to qubit $A$ \cite{acin00}, that reads as
\begin{equation}
\Psi=a\left\vert 000\right\rangle +be^{i\varphi}\left\vert 100\right\rangle
+g\left\vert \Phi_{1}\right\rangle , \label{state}%
\end{equation}
where $g=\sqrt{c^{2}+d^{2}+f^{2}}$ and $\Phi_{1}=\frac{c}{g}\left\vert
110\right\rangle +\frac{d}{g}\left\vert 101\right\rangle +\frac{f}%
{g}\left\vert 111\right\rangle $. The partial transpose of the state operator
$\widehat{\rho}=$ $\left\vert \Psi\right\rangle \left\langle \Psi\right\vert $
for subsystem $A$ in the vector space spanned by basis vectors $\left\vert
000\right\rangle $, $\left\vert 100\right\rangle $, $\left\vert \Phi
_{1}\right\rangle $ and $\left\vert \Phi_{1}^{T}\right\rangle =\frac{c}%
{g}\left\vert 010\right\rangle +\frac{d}{g}\left\vert 001\right\rangle
+\frac{f}{g}\left\vert 011\right\rangle ,$ is%
\begin{equation}
\rho_{G}^{T_{A}}=\left[
\begin{array}
[c]{cccc}%
a^{2} & abe^{i\varphi} & 0 & 0\\
abe^{-i\varphi} & b^{2} & be^{-i\varphi}g & ag\\
0 & be^{i\varphi}g & g^{2} & 0\\
0 & ag & 0 & 0
\end{array}
\right]  . \label{rotg}%
\end{equation}
The partially transposed matrix $\rho_{G}^{T_{A}}$ is a negative matrix with a
negative eigenvalue $\lambda^{-}=-ag$ and the corresponding eigenvector%
\begin{equation}
\left\vert \Psi_{G}^{-}\right\rangle =\frac{\left(  a+g\right)  }{\sqrt
{4ag+2}}\left\vert 100\right\rangle -\frac{be^{i\varphi}}{\sqrt{4ag+2}%
}\left\vert 000\right\rangle -\frac{be^{i\varphi}}{\sqrt{4ag+2}}\left\vert
\Phi_{1}\right\rangle -\frac{\left(  a+g\right)  }{\sqrt{4ag+2}}\left\vert
\Phi_{1}^{T}\right\rangle \text{.} \label{eig}%
\end{equation}
It is easily verified that
\begin{equation}
\rho_{G}^{T_{A}}=\widehat{\rho}_{3}^{T_{A}}+\widehat{\rho}_{2}^{T_{A}%
}-\widehat{\rho}. \label{d}%
\end{equation}
Using the relation given by Eq. (\ref{d}) in%
\begin{equation}
N_{G}^{A}=-2\left\langle \Psi_{G}^{-}\right\vert \rho_{G}^{T_{A}}\left\vert
\Psi_{G}^{-}\right\rangle =2ag, \label{neg}%
\end{equation}
we obtain%
\begin{equation}
N_{G}^{A}=E_{3}^{A}+E_{2}^{A}-E_{0}^{A},
\end{equation}
where the partial $K-$way negativities $E_{K}^{A}$ for $K=2$ and $3$ are
defined as
\begin{equation}
E_{K}^{A}=-2\left\langle \Psi_{G}^{-}\right\vert \widehat{\rho}_{K}^{T_{A}%
}\left\vert \Psi_{G}^{-}\right\rangle ,\qquad\text{while }\qquad E_{0}%
^{A}=-2\left\langle \Psi_{G}^{-}\right\vert \widehat{\rho}\left\vert \Psi
_{G}^{-}\right\rangle .
\end{equation}
We may further write
\begin{equation}
\rho_{2}^{T_{A}}=\widehat{\rho}_{2}^{T_{A-AB}}+\widehat{\rho}_{2}^{T_{A-AC}%
}-\widehat{\rho},
\end{equation}
and define
\begin{equation}
E_{2}^{A-AB}=-2\left\langle \Psi_{G}^{-}\right\vert \widehat{\rho}%
_{2}^{T_{A-AB}}\left\vert \Psi_{G}^{-}\right\rangle ,\qquad E_{2}%
^{A-AC}=-2\left\langle \Psi_{G}^{-}\right\vert \widehat{\rho}_{2}^{T_{A-AC}%
}\left\vert \Psi_{G}^{-}\right\rangle ,
\end{equation}
giving $E_{2}^{A}=E_{2}^{A-AB}+E_{2}^{A-AC}.$ For the state of Eq. (
\ref{eig}), we obtain%
\begin{equation}
E_{3}^{A}=\frac{4a^{2}f^{2}}{2ag},\qquad E_{2}^{A}=\frac{4a^{2}\left(
c^{2}+d^{2}\right)  }{2ag}, \label{ep3}%
\end{equation}
leading to%
\begin{equation}
\left(  N_{G}^{A}\right)  ^{2}=E_{3}^{A}N_{G}^{A}+E_{2}^{A}N_{G}^{A}.
\label{negsq}%
\end{equation}
The negative part of $\rho_{G}^{T_{A}}$ is found to be orthogonal to
$\widehat{\rho}$ giving $E_{0}^{A}=0$. We also obtain
\begin{equation}
E_{2}^{A-AB}=\frac{4a^{2}c^{2}}{N_{G}^{A}},\qquad\text{and\qquad}E_{2}%
^{A-AC}=\frac{4a^{2}d^{2}}{N_{G}^{A}}.
\end{equation}

The tangle, defined to be a measure of entanglement of qubit $A$ with
subsystem $BC$ \cite{woot98}, for the pure state of Eq. ( \ref{eig}) is given
by
\begin{align*}
\tau_{A(BC)}  &  =4\det\left(  \rho^{A}\right)  =4\det\left[
\begin{array}
[c]{cc}%
a^{2} & be^{i\varphi}\\
abe^{-i\varphi} & \left\vert b\right\vert ^{2}+\left\vert g\right\vert ^{2}%
\end{array}
\right] \\
&  =4a^{2}g^{2}.
\end{align*}
If tangles $\tau_{AB}$ and $\tau_{AC}$ are measures of entanglement of $A$
with $B$ and $C$ respectively, the three tangle \cite{coff00} defined as%
\begin{equation}
\tau_{3}=\tau_{A(BC)}-\tau_{AB}-\tau_{AC}, \label{3tangle}%
\end{equation}
is known to measure the tripartite entanglement of the pure state. The value
of $\tau_{AB}$ for the mixed states $\rho^{AB}=tr_{C}(\rho^{ABC})$ is obtained
by constructing the spin flipped density matrix%
\[
\widetilde{\rho^{AB}}=\left(  \sigma_{y}\otimes\sigma_{y}\right)  \left(
\rho^{AB}\right)  ^{\ast}\left(  \sigma_{y}\otimes\sigma_{y}\right)  ,
\]
where asterisk denotes complex conjugation and $\sigma_{y}=\left[
\begin{array}
[c]{cc}%
0 & -i\\
i & 0
\end{array}
\right]  $. With the square roots of the eigenvalues of product matrix
$\rho^{AB}\widetilde{\rho^{AB}}$ given by $\lambda_{1}$, $\lambda_{2}$,
$\lambda_{3}$, and $\lambda_{4}$, the tangle $\tau_{AB}$ is defined
\cite{woot98} as $\tau_{AB}=\left[  \max\left\{  \lambda_{1}-\lambda
_{2}-\lambda_{3}-\lambda_{4},0\right\}  \right]  ^{2}.$ For the generic state
of Eq. ( \ref{eig}) \ we obtain $\tau_{AB}=4a^{2}c^{2}$. The tangle $\tau
_{AC}$ calculated for $\rho^{AC}=tr_{B}(\rho^{ABC})$ has the value
$4a^{2}d^{2}$. As such, the three tangle, for the generic three qubit
canonical state of Eq. (\ref{state}), is found to be $\tau_{3}=4a^{2}f^{2}$
giving%
\begin{equation}
E_{3}^{A}N_{G}^{A}=\tau_{3},\qquad E_{2}^{A}N_{G}^{A}=\tau_{AB}+\tau
_{AC},\qquad\left(  N_{G}^{A}\right)  ^{2}=\tau_{A(BC)}.
\end{equation}
We also identify the products $E_{2}^{A-AB}N_{G}^{A}$ and $E_{2}^{A-AC}%
N_{G}^{A}$ with tangles $\tau_{AB}$ and $\tau_{AC}$, respectively. The product
$E_{K}^{A}N_{G}^{A}$ measures the $K-$way ($K=2$ and $3$) entanglement of the
canonical state. For a state obtained from the canonical state by local
unitary transformations on qubits $A$, $B$, and $C$, the partial $3-$way
negativity may be different from $E_{3}^{p}$, $p=1$ to $3$, but the global
negativity remains invariant. As such, for a given three qubit state, the
difference $\Delta=$ $E_{3}^{A}N_{G}^{A}-\tau_{3}$ is a measure of the amount
of two-qubit coherences transformed to three qubit coherences by applying
local unitary transformations on the corresponding canonical state.

To illustrate, how local operations on qubits transform three qubit coherences
into two qubit coherences and vice versa, we consider a single parameter
GHZ-like state
\begin{equation}
\Psi_{a}=a\left\vert 000\right\rangle +\sqrt{1-a^{2}}\left\vert
111\right\rangle ,\qquad0<a<1, \label{GHZ2}%
\end{equation}
with parties $A$, $B$ and $C$ holding one qubit each. The global negativity
$N_{G}^{A}=2a\sqrt{1-a^{2}}$ is equal to $E_{3}^{A}$, while $E_{2}^{A}=0.$ We
define a unitary rotation of qubit $C$ by the transformation matrix
\begin{equation}
U^{C}=\left[
\begin{array}
[c]{cc}%
\cos\left(  \frac{\alpha}{2}\right)  & \sin\left(  \frac{\alpha}{2}\right) \\
-\sin\left(  \frac{\alpha}{2}\right)  & \cos\left(  \frac{\alpha}{2}\right)
\end{array}
\right]  .
\end{equation}
For the state $U^{C}\left\vert \Psi_{a}\right\rangle $ we obtain%
\begin{align}
E_{3}^{A}\left(  \alpha\right)   &  =\frac{a\sqrt{1-a^{2}}}{2}\left(
3+\cos\left(  2\alpha\right)  \right)  ,\nonumber\\
E_{2}^{A}\left(  \alpha\right)   &  =\frac{a\sqrt{1-a^{2}}}{2}\left(
1-\cos\left(  2\alpha\right)  \right)  ,
\end{align}
showing that the maximum amount of three way correlations in the state
$\Psi_{a}$ that may be transformed to two-qubit coherences is $N_{G}^{A}%
E_{2}^{A}\left(  \pi/2\right)  =$ $\left(  N_{G}^{A}\right)  ^{2}/2$. We
notice that a state measurement, in logical basis, on qubit $C$ completely
destroys the entanglment of state $\Psi_{a}$, while in the case of
$U^{C}\left\vert \Psi_{a}\right\rangle $ destroys entanglement due only to
coherences measured by $E_{3}^{A}\left(  \alpha\right)  $. Consequently, the
reduced state tr$_{C}\left(  \left\vert \Psi_{a}\right\rangle \left\langle
\Psi_{a}\right\vert \right)  $ is a sepatable state, while tr$_{C}\left(
U^{C}\left\vert \Psi_{a}\right\rangle \left\langle \Psi_{a}\right\vert \left(
U^{C}\right)  ^{\dagger}\right)  $ is an entangled two-qubit mixed state.

For three qubit states in canonical form, GHZ-like states have $N_{G}^{A}=$
$E_{3}^{A}$, while W-like states have $N_{G}^{A}=$ $E_{2}^{A}$. The N-qubit
states, may, likewise be classified according to the amount and type of
$K-$way coherences ($2\leq K\leq N$) present in the canonical form. It was
shown in ref. \cite{shar08} that for an N partite system, the global
negativity of partially transposed state operator $\widehat{\rho}$ with
respect to a subsystem $p$ can be written in terms of partial $K-$way
negativities as
\begin{equation}
N_{G}^{p}=\sum\limits_{K=2}^{N}E_{K}^{p}-E_{0}^{p},
\end{equation}
where%
\begin{equation}
E_{K}^{p}=-\frac{2}{d_{p}-1}\sum\limits_{i}\left\langle \Psi_{i}%
^{G-}\right\vert \widehat{\rho}_{K}^{T_{p}}\left\vert \Psi_{i}^{G-}%
\right\rangle ,\qquad2\leq K\leq N,
\end{equation}%
\begin{equation}
E_{0}^{p}=-\frac{2(N-2)}{d_{p}-1}\sum\limits_{K=2}^{N}\sum\limits_{i}%
\left\langle \Psi_{i}^{G-}\right\vert \widehat{\rho}\left\vert \Psi_{i}%
^{G-}\right\rangle ,
\end{equation}
and $\Psi_{i}^{G-}$ the eigenvector of $\widehat{\rho}_{G}^{T_{p}}$
corresponding to negative eigenvalue $\lambda_{i}$. For the case where
$E_{0}^{p}=0$, we can affirm that
\begin{equation}
N_{G}^{p}\geq E_{K}^{p},\qquad2\leq K\leq N.
\end{equation}
It is important to note that if $N_{G}^{p}>0$ for $p$ referring to a part in
all possible bipartite splits of the composite quantum system, then $E_{N}%
^{p}=0$ for a given $p$ does not mean that N-partite entanglement is zero. The
N-partite entanglement in this case could be caused by coherences of the order
less than N. For the special case of three qubit states, we obtain two
inequalities
\begin{equation}
N_{G}^{p}\geq E_{2}^{p},\qquad N_{G}^{p}\geq E_{3}^{p},
\end{equation}
the former being equivalent to the CKW inequality \cite{coff00}%
\begin{equation}
\tau_{A(BC)}\geq\tau_{AB}+\tau_{AC},
\end{equation}
on which the definition of three tangle is based.

\section{A linear combination of three qubit GHZ state and W state}

An interesting pure state for which tripartite entanglement has been studied
by Lohmayer et al. \cite{lohm06} \ is a superposition of three qubit GHZ state
and W state. In our paper \cite{shar07}, we calculated $N_{G}^{A},E_{3}^{A},$
and $E_{2}^{A}$ for single parameter pure states
\begin{equation}
\Psi^{(\pm)}=\sqrt{q}\Psi_{GHZ}\pm\sqrt{(1-q)}\Psi_{W},\quad\widehat{\rho}%
_{2}^{\pm}=\left\vert \Psi_{2}^{(\pm)}\right\rangle \left\langle \Psi
_{2}^{(\pm)}\right\vert \qquad0\leq q\leq1\label{22}%
\end{equation}
and compared with three tangle for states $\Psi^{(\pm)}$, calculated from%
\begin{equation}
\tau_{3}\left(  \Psi^{(\pm)}\right)  =\left\vert q^{2}\pm\frac{8\sqrt
{6(q(1-q)^{3}}}{9}\right\vert .\label{qtan}%
\end{equation}
Here $\Psi_{GHZ}=\left(  \left\vert 000\right\rangle +\left\vert
111\right\rangle \right)  /\sqrt{2}$, is GHZ state for which $\tau_{3}\left(
\Psi_{GHZ}\right)  =1$ and%
\[
\Psi_{W}=\frac{\left\vert 100\right\rangle +\left\vert 010\right\rangle
+\left\vert 001\right\rangle }{\sqrt{3}},
\]
is W state with $\tau_{3}\left(  \Psi_{W}\right)  =0$. It was found that both
for the state $\Psi^{(+)}$ and $\Psi^{(-)\text{ }}$the partial $3-$way
negativity $E_{3}^{A}>0$ for parameter values $0<q<1$. To put the record
straight, the states $\Psi^{(\pm)}$ are not in canonical form. Therefore the
difference $\Delta=$ $E_{3}^{A}N_{G}^{A}-\tau_{3}$ for a given state, observed
in \cite{shar07} measures the distance of the state from the canonical state
in terms of excess or deficit of three qubit coherences in comparison with
that for the canonical state. The states $\Psi^{(\pm)}$ can be transformed to
Schmidt like form for qubit $A,$ by following the procedure given by Acin et
al. \cite{acin00}. A unitary transformation%
\begin{equation}
U^{A}=\left[
\begin{array}
[c]{cc}%
\alpha & -\beta\\
\beta & \alpha
\end{array}
\right]  ,\qquad\left\vert \alpha\right\vert ^{2}+\left\vert \beta\right\vert
^{2}=1\label{UA}%
\end{equation}
on qubit $A$, followed by Unitary transformations
\begin{equation}
U^{C}=\frac{1}{\sqrt{a^{2}\beta^{2}+b^{2}\alpha^{2}}}\left[
\begin{array}
[c]{cc}%
\beta a & \alpha b\\
-\alpha b & \beta a
\end{array}
\right]  ,\qquad a=\sqrt{\frac{q}{2}},\qquad b=\pm\sqrt{\frac{1-q}{3}%
,}\label{UB}%
\end{equation}%
\begin{equation}
U^{B}=\frac{1}{\sqrt{\left(  a^{2}\beta^{2}+b^{2}\alpha^{2}\right)  }}\left[
\begin{array}
[c]{cc}%
-b\alpha & a\beta\\
-a\beta & -b\alpha
\end{array}
\right]  ,\label{UC}%
\end{equation}
on qubits $C$ and $B$, subject to the constraint $\alpha\beta a^{2}-\beta
^{2}ab+\alpha^{2}b^{2}=0,$yield the state%
\[
\Psi_{c}(q)=a_{000}\left\vert 000\right\rangle +a_{100}\left\vert
100\right\rangle +a_{110}\left\vert 110\right\rangle +a_{101}\left\vert
101\right\rangle +a_{111}\left\vert 111\right\rangle .
\]
The two possible solutions are
\[
\alpha=\beta x^{2}\frac{1}{2}\left(  1\pm\sqrt{1-\frac{4}{x^{3}}}\right)
,\qquad x=\frac{a}{b}=\mp\sqrt{\frac{3q}{2(1-q)}},
\]
except\ for the case $x^{3}=4$ ( $q=0.62685$ for state $\Psi^{(-)}),$ in which
case only solution is $\alpha=\beta x^{2}/2$, giving $\alpha=0.783\,27$ and
$\beta=0.621\,69.$ For $x^{3}\neq4$, the partial $3-$way negativity of the
state (using Eq.$\left(  \text{\ref{ep3}}\right)  ),$
\begin{equation}
E_{3}^{A}=\frac{2a_{000}a_{111}^{2}}{\sqrt{1-\left(  a_{000}\right)
^{2}-\left(  a_{111}\right)  ^{2}}},
\end{equation}
is non zero for $a_{000}\neq0$, and $a_{111}\neq0$.

The canonical state for $x^{3}=4$ reads as
\begin{equation}
\Psi_{c}(q=0.62685)=-0.567\,31\left\vert 010\right\rangle -0.567\,31\left\vert
100\right\rangle +0.185\,78\left\vert 110\right\rangle +0.567\,31\left\vert
111\right\rangle \text{,} \label{ep3zero}%
\end{equation}
giving $N_{G}^{A}=E_{2}^{A}=0.910\,3,\qquad E_{3}^{A}=0.$This result is
consistent with the fact that three tangle $\tau_{3}\left(  \Psi^{-)}\right)
=0$ for $q=0.62685.$ The W-like tripartite entanglement of the state
\ $\Psi_{c}(q=0.62685)$ arises solely from two-body coherences measured by
nonzero $E_{2}^{A}.$of the canonical state. The GHZ like tripartite
entanglement of a three qubit state is due to three-body coherences (nonzero
$E_{3}^{A}$ of canonical state). Two- body correlations generate W-like
tripartite entanglement and pairwise entanglement of qubits. A mixed state,
$\widehat{\rho}_{mixed},$ can be decomposed as a convex combination of
projectors onto pure states as%
\[
\widehat{\rho}_{mixed}=\sum_{i}P_{i}\widehat{\rho}_{i},\qquad\widehat{\rho
}_{i}=\left\vert \Psi_{i}\right\rangle \left\langle \Psi_{i}\right\vert ,
\]
where $P_{i}$ is the probability of fnding the state $\widehat{\rho}_{i}.$ For
the mixed states, we define the convex roof extended Global and $K-$way
negativities through
\begin{equation}
N^{p}(\widehat{\rho}_{mixed})=\min\sum\limits_{i}P_{i}N^{p}\left(
\widehat{\rho}_{i}\right)  .
\end{equation}
Here minimum of $\sum\limits_{i}P_{i}N^{p}\left(  \widehat{\rho}_{i}\right)  $
taken over all possible decompositions of $\widehat{\rho}_{mixed}$ is defined
as $N^{p}(\widehat{\rho}_{mixed})$. The negativities $\left(  N_{G}%
^{A}(\widehat{\rho}_{mixed}^{AB})\right)  ^{2}$ and $\left(  N_{G}%
^{A}(\widehat{\rho}_{mixed}^{AC})\right)  ^{2}$ , where $\widehat{\rho
}_{mixed}^{AB}=tr_{C}(\widehat{\rho})$ and $\widehat{\rho}_{mixed}^{AC}%
=tr_{B}(\widehat{\rho})$, measures the pairwise entanglement in three qubit
state $\widehat{\rho}.$ For the state of Eq. (\ref{ep3zero}), we get%
\begin{equation}
\left(  N_{G}^{A}(\widehat{\rho}_{mixed}^{AB})\right)  ^{2}=\left(  N_{G}%
^{A}(\widehat{\rho}_{mixed}^{AC})\right)  ^{2}=0.414\,3
\end{equation}
that is%
\begin{equation}
\left(  N_{G}^{A}\right)  ^{2}=\left(  N_{G}^{A}(\widehat{\rho}_{mixed}%
^{AB})\right)  ^{2}+\left(  N_{G}^{A}(\widehat{\rho}_{mixed}^{AC})\right)
^{2}=0.828\,6.
\end{equation}
This analysis leads to a simple explanation for why a GHZ state can not be
converted to a W-state. It is not possible to transform three-body
correlations into purely two-body correlations by local operations and
classical communication.

To summarize, we have shown that for generic three qubit canonical state the
product of global negativity and partial three way negativity is equal to
three tangle, which is an entanglement monotone, quantifying GHZ-like three
qubit quantum correlations. The product of global negativity and partial two
way negativity for a given pair of qubits in the canonical state is seen to be
equal to tangle for the pair. It proves that the $K-$way negativities
associated with the three qubit canonical state are indeed entanglement
measures. The importance of this result lies in the ease with which the
partial $K-$way negativities can be calculated for multipartite canonical
states and the physical meaning associated to partial $K-$way negativities as
measures of $K-$partite coherences. The global negativity and partial $K-$way
negativities, obtained by selective partial transpositions on multi-qubit
state operator, satisfy inequalities which for three qubits are equivalent to
CKW (Coffman-Kundu-Wootter) inequality. The difference between the values of
product of global and partial three way negativity for a given state and three
tangle for the state is a quantitative measure of two qubit correlations
transformed by unitary transformations on canonical state into three qubit
correlations. We have also calculated the partial $K-$way negativities and
three tangle for the state canonical to a single parameter ($0<q<1$) pure
state which is a linear combination of a GHZ state and a W state. In this case
for $q=0.62685,$ the state has zero three tangle as well as zero three-way
negativity, having only W-like entanglement. For mixed states the relevant
entanglement measures are convex roof extended global and $K-$way
negativities. The GHZ-like and W-like state are found to differ in the amount
and type of quantum correlations present in canonical form. The N-qubit states
may also be classified according to the amount and type of $K-$way coherences
($2\leq K\leq N$) present in the canonical form. We expect the use of partial
$K-$way negativities to enhance our understanding of quantum correlations and
entanglement in multipartite systems.

Financial support from Faep-UEL, Brazil is acknowledged.

\end{document}